# A Different Perspective on Retirement Income Sustainability:
## The Blueprint for a Ruin Contingent Life Annuity (RCLA)

By[1]: H. Huang, M. A. Milevsky and T.S. Salisbury
York University, Toronto, Canada

*Version: 18 July 2008*

**ABSTRACT:**

*The purpose of this article is twofold. First, we motivate the need for a new type of stand-alone retirement income insurance product that would help individuals protect against personal longevity risk and possible "retirement ruin" in an economically efficient manner. We label this product a ruin-contingent life annuity (RCLA), which we elaborate-on and explain with various numerical examples and a basic pricing model. Second, we argue that with the proper perspective a similar product actually exists, albeit not available on a stand-alone basis. Namely, they are fused and embedded within modern variable annuity (VA) policies with guaranteed living income benefit (GLiB) riders. Indeed, the popularity of GLiB riders on VA policies point towards the potential commercial success of such a stand-alone vehicle.*

**INTRODUCTION:**

The *raison d'être* of the variable annuity (VA) industry – now with over $1.4 trillion in assets and over $160 billion in annual sales[2] -- has shifted from tax deferral and death benefits to income riders. The large majority of VA sales – a.k.a., Segregated funds in Canada -- now include guaranteed living withdrawal benefit riders, which anecdotally have become central to the sales pitch and allure. To the insurance companies manufacturing the new generation of variable annuities, they are viewed as an ideal private sector replacement for defined benefit (DB) pensions, in an increasingly defined contribution (DC) world.

[1] Moshe A. Milevsky, the contact author, can be reached at tel: 416-736-2100 or via email at milevsky@yorku.ca. Huang and Salisbury would like to acknowledge funding from NSERC and MITACS and all three authors would like to acknowledge the resources of The IFID Centre at the Fields Institute for Research in Mathematics. The authors would like to acknowledge the helpful comments of an anonymous reviewer for the Journal of Wealth Management, as well as the editor Jean Brunel and associate editor Bill Reichenstein.
[2] Source: NAVA Quarterly Report, June 15, 2007

Motivated by the popularity of these products – and the fundamental fear they appear to help address -- in this concept article we pose the question: why must retirees buy this type of insurance <u>with</u> their investments? Would it not be possible to purchase the same insurance <u>for</u> their investments and keep the insurance coverage distinct from the money management processes? In other words, we are proposing that a new insurance product be developed that effectively bifurcates or separates the variable annuity into distinct components: insurance product vs. investment product.

This might sound confusing at first – and academic at worst -- but it truly gets to the heart of the retirement income dilemma. Currently, the variable annuity "story" or bundle contains two distinct parts. The first component is the promise of a series of cash flows -- which is essentially a reverse dollar cost averaging strategy -- that continues until the day and time the investment account hits zero. The second component begins if-and-when the account is depleted or ruined, at which point the embedded insurance comes into existence. A second series of payments continue from a depletion or ruin date, and these cash flows continue for a single or dual life; which is why VAs are also positioned as providing longevity insurance for DB-deprived retirees. This is true regardless of the exact mechanics of the rider, whether it is a GMIB or GMWB, etc.

*But why must these two services be bundled and offered in the same product and by the same provider? Alternatively, even if this is offered by one provider, why must consumers purchase both instruments together?*

In terms of background, a number of recent[3] articles have explored the important role of pure longevity insurance, i.e., a deferred life annuity that initiates payments well into the retirement years, e.g. age 80 or 85, as opposed to at the start of retirement. A collection of related products now fall under the umbrella of advanced life delayed annuities (ALDA), a.k.a. longevity insurance with a deductible. Many U.S. companies such as New York Life, Hartford Life, MetLife and Prudential Financial are now offering variants

---

[3] See for example the paper by Milevsky (2005), Stephenson (1978), Webb Gong and Sun (2007)

on this concept. A recent article in the popular and widely-read *Money Magazine* in January 2008 is yet another indication of the public's interest in longevity annuities.

Our main objective in this paper is take the delayed annuity concept one (final) step further by fusing its desirable longevity insurance and inflation hedge concept with actual investment portfolio insurance. Stated differently, we propose an ALDA in which two distinct risk valves must be triggered before the annuitant gets paid. First and obviously the individual (or their spouse) must be alive. But second, to make a claim on their insurance policy, they must have the ill fortune to experience under-average market returns during the years surrounding retirement. If markets produce above-average returns, the insurance policy will not pay off. Stated more crudely, if you happen to get "lucky in life" then why bother with a life annuity? After all, Bill Gates doesn't need longevity insurance.

Note that the product we propose is already embedded deeply and obscurely within many guaranteed living income benefits (GLiBs) offered as riders on variable annuities. In some sense, this article can also be viewed as providing a better understanding of the precise nature of the insurance offered by the latest generation of variable annuities.

To flesh out this concept, in the next few pages we describe how such a product would work, a rudimentary pricing methodology, and then return to justifying the motivation and need for what we are calling a ruin-contingent life annuity (RCLA.) And, although the acronym might sound clunky, we are emphasizing the ruin-contingent aspect of the payout. This is a form of contingent annuitization; only if markets perform poorly will the annuitization kick-in.

Even if retirees are not necessarily interested in actually purchasing the RCLA to hedge their retirement risks, the mere existence of such products -- and hopefully their transparent prices -- would provide individuals with a mark-to-market value of their retirement spending and consumption strategies. In particular, many retirees currently buy products such as variable annuities that bundle together many components - mutual funds, death benefits, income benefits, etc. Fees are charged on the whole package, and it is a complex question for both retirees and financial advisors to disentangle the components. If component prices (eg. RCLAs) are available from the market directly, then both retirees and financial advisors will be able to decide more easily which products provide the desired protection at a reasonable price. We will follow-up on this aspect of the proposal in the concluding section of this article.

## HOW DOES A GMWB WORK?

We start, for comparative purposes, by describing a product which already exists in the market, a *Guaranteed Minimum Withdrawal Benefit* (GMWB). Variable annuities are a widely used vehicle for retirement savings, and a large fraction of them are now sold with GMWB riders, or other similar guaranteed living benefits (GLiB's). Variable annuities became popular initially because of the favorable tax treatment they enjoy. But as these products have moved to include guarantees on the minimum income stream investors would receive, those features have become important selling points. Essentially a GMWB rider allows investors to lock in income for life, without tying up or surrendering their capital. Thus they provide the downside protection of a traditional annuity, without surrendering all upside potential or liquidity.

GMWBs exist in a variety of formats, and are often bundled with an array of other guarantees, ratchets, or stepups. So for simplicity we will consider an idealized version. For us, the underlying variable annuity consists of an investment account (or mutual fund), from which one makes periodic withdrawals. For example 5% per year of the initially deposited amount. The GMWB guarantee is that these withdrawals will continue for life, regardless of whether the underlying account has the funds to support them. In other words, fees and withdrawals are deducted from the variable annuity account, as long as there are funds available there. But if those periodic withdrawals ever fully deplete this account, the underwriter steps in and funds the remaining withdrawals, for the lifetime of the investor.

The periodic withdrawals provide downside protection, but there is still upside potential for the underlying account to grow if markets perform well. The investor preserves liquidity, since the underlying account value may be withdrawn at any time (less any surrender charges). Unlike a traditional annuity, if the investor dies, his or her heirs will inherit the remaining account value. When the associated fees are reasonable, this blend of features has proved popular with investors. In this paper we take this development one step further, and think of the sequence of GMWB cash flows as coming in two parts -- the withdrawals prior to the account value hitting zero (the ruin time), and the cash flows after this ruin time (if indeed ruin ever occurs). Clearly the first series is most naturally viewed as an investment product, while the second is really an insurance product. This second series of cash flows is what we wish to focus on, and will advocate making available as a stand-alone product.

**HOW DOES AN RCLA WORK?**

The basic concept underlying the ruin-contingent life annuity (RCLA) is that it provides insurance against the joint occurrence of two separate and independents events; namely poor market returns, especially during the first few years of retirement spending (a.k.a., the retirement risk zone) and above average or unplanned survival rates. It is a life annuity that only kicks-in under poor economic scenarios. This should appeal to the many individuals who would like to use annuitization only as a worst case scenario.

Like any derivative product, the RCLA requires an underlying security or economic state variable upon which payments would be triggered and calculated. Thus, the first ingredient or main component of the RCLA would be a new and unique pseudo-equity index that would track the total return of the SP500 (for example, although it could be the Russell or the FTSE) over time: but subjected to periodic withdrawals. We would label this number the SwP index to evoke the notion of a systematic withdrawal plan that is linked to the SP500 index, although once again any diversified equity or balanced index could be used.

Technically speaking, the SwP pseudo-index would start with a vintage year, e.g. 2007, would then be adjusted by the total returns of its constituent stocks, but would also be reduced by a withdrawal factor, which mimics portfolio withdrawals. Over time the SwP pseudo-index would be expected to decline, quite similar to the anticipated behavior of a retirement nest egg or portfolio. This SwP pseudo-index would of course depend on a pre-specified withdrawal rate or factor, which would be fixed and determined at the launch of the pseudo-index.

Here is a detailed example that should help explain the pseudo-index. Imagine the SP500 index was hypothetically at a level of 1000 on January $1^{st}$, 2007. Under a pre-specified withdrawal factor rate of 7% -- assuming that during January 2007 the consumer price index increased by 0.5%, while the SP500 return was a nominal 2% -- then the level of a vintage 2007 SwP pseudo-index on the first day of February 2007 would be $100 x (1+0.02) – (7/12) x (1+0.005) = $101.94. The annual withdrawal rate of $7 is divided by 12 to create the monthly withdrawal, which is adjusted for inflation. The same calculation algorithm would continue each month. The calculation would gross-up by total return including dividends, subtract monthly withdrawals adjusted for inflation and generate the new level.

Now, if and when this vintage-2007 SwP 7% pseudo-index ever hit zero, the insurance company would then commence making payments to the annuitant who bought the product in early 2007, as long as they were still alive. The exact amount of the payment would be 7% of the original $100 value of the index, which is $7 adjusted for inflation per year, for life.

Naturally, the individual could scale up and purchase 1000, 2000 or 5000 such RCLA units to protect a $100,000 or $200,000 or $500,000 nest egg. Likewise, the retiree could select from a range of withdrawal rates, for example 4%, 6% or 8%, assuming the insurance company was willing to offer a menu of spending rates (at different prices, of course.) In fact, the underlying index itself could be a combination of the SP500, the Russell 3000 or even the MSCI index. They are just variations on a theme.

| FIGURE #1 PLACED HERE |
| --- |

Here is a graphical illustration of how the SwP pseudo-index has evolved over time. For example – as you can see from Figure #1 -- had you purchased a vintage 1970 SP500-based *RCLA struck at 7%*, then your lifetime income payments would have started in January of 1983. This is a mere 13 years after your so-called retirement date. The size of the payment would depend on the number of RCLA units you acquired.

In contrast to the 1970 vintage, had you purchased the same *RCLA struck at 7%* in January of 1976, which is 6 years later, you would actually still be waiting for payments to commence thirty years later in January 2006. This is because the vintage 1976 SwP pseudo-index has yet to hit zero. In fact, it is "trading" at a level of $400 in early 2007, and so it is safe to say that vintage 1976 will never pay-off. Basically, the market returns during the first 30 years have been strong enough to overcome inflation-adjusted withdrawals of 7%.

But then again, the vintage 1973 RCLA struck at 7% would have kicked in (paid off) exactly 10 years later in January 1983. As most practitioners are well aware, the sustainability of a given spending rate is extremely sensitive to the initial sequence of returns during the first few years of retirement.

Of course, we remind the reader that in all these cases the retiree would have to be alive on the precise date of pseudo-index "ruin" for the RCLA annuitant to start receiving payments. Moreover, the retiree would have to stay alive to continue getting these payments. But, this is just like any other life contingent annuity.

| TABLE #1 PLACED HERE |
| --- |

As one can see from Table #1 – which is a true historical analysis -- whether or not the RCLA ever pays out (or "matures in the money", in the language of options) is quite sensitive to both the spending rate as well as the vintage year in which the pseudo-index is initiated. Notice that a hypothetical RCLA struck at 4% would have never paid out regardless of whether the vintage date was 1970, 1973, 1976 or 1979. In contract, an RCLA struck at 9% would have resulted in a payout for three of the four starting dates.

In practice – to keep this process manageable and intuitive -- we envision an insurance company creating and updating a family of SwP pseudo-indices which are struck under a limited number of spending rates with ongoing issuance of annual vintages depending on demand. A variety of underlying equity indices would be available as well.

To start out, we recommend a 4%, 6% and 8% SwP pseudo-index that would launch January 2009 at a hypothetical value of $1,000. Each month this value would be increased by the total return of the SP500 index, but then reduced by the relevant CPI-adjusted spending amount of $40/12 or $60/12 or $80/12. These three pseudo-index levels would be calculated and published in early February 2009, etc.

Intuitively, the vintage 2009 SwP pseudo-index struck at 4% would always be slightly higher than the one struck at 6%, which would be slightly higher than the one struck at 8%, due to the lower withdrawals. Then, in January 2010 a new vintage of SwP pseudo-indices would be created for a hypothetical retiree starting withdrawals in January 2010, etc. Over time this might accumulate to (quite) a large matrix of indices, but this should be no more cumbersome than keeping track of a changing yield curve or swap curve over long maturities. In fact, the recent popularity of target-date or retirement-date lifecycle funds prove that it is possible to keep track of many different vintage-year fund classes that differ by just a few equity percentage points.

A retiree with a diversified portfolio of international stocks or perhaps high-yield bonds might buy various types of RCLA, each linked to different underlying indices so as to best hedge their particular nest egg. Once again, these are implementation details. In fact, the point is not necessarily to create a perfect hedge to a fixed fictional spending rate, but rather a rough insurance policy against these broad risks.

## WHY WOULD ANYBODY BUY THIS?

As we argued above, there seems to be a consensus[4] amongst researchers in the field of retirement income planning that when individuals transition into their portfolio retirement years they face a number of unique risks. These risks are related to uncertain inflation, poor investment returns especially early-on in retirement and unexpected longevity. In fact, these three risks are often referenced as the main reason why retirees should consider purchasing a variable annuity (VA) with guaranteed living benefit (GLiBs). Even those with substantial income from (private sector) defined benefit pension plans are un-hedged or un-protected against inflation risks. Only Federal Social Security payments and pension income from military pensions are completely insulted from inflation risks.

Yet, if a bear market at the "wrong time" has such a substantial impact on retirement income sustainability, then why not buy insurance against these risks without having to give up or cede control of investment assets? Why does the insurance company have to take control of the assets in order to provide the insured with the desired protection? In theory, the retiree could continue to manage his or her own portfolio -- maintaining liquidity and withdrawing more/less as needed over time -- knowing that if they do experience an adverse economic shock, they will continue to receive lifetime income. Recall that longevity risk is only costly to the retiree within the states of nature in which the investment portfolio has earned less than expected. After all, if a retiree with $100,000 nest egg is withdrawing (a normally unsustainable) $7,000 per year in inflation-adjusted terms and by sheer luck the SP500 earns 20% per year over the next 5 years, longevity risk is no longer an issue. *Ex post*, the retirement portfolio can now sustain much greater withdrawals. Clearly then, longevity risk is intertwined with

---

[4] See for example the book edited by Evensky and Katz (2006).

portfolio risk and stand-alone ALDAs – or any fixed immediate annuities for that matter -
- do not hedge the retiree's combined risks.

## HOW MUCH WOULD IT COST?

It is obviously quite difficult to predict how much an insurance company would charge
for an RCLA, since it depends on many factors that are difficult to quantify in advance.
Intuitively though, it would depend on the age of the buyer since payments are due
upon ruin only if the annuitant is still alive. Thus, the cost of an RCLA struck at x%
would be lower, the higher the age of the buyer. Likewise, all else being equal, a 5%
RCLA would be cheaper than a 7% RCLA which in turn would be cheaper than a 10%
RCLA, etc. Intuitively, the higher the spending rate, the higher is the chance – and the
sooner is the date -- the insurance will kick-in. Also, although we envision the RCLA as
being linked to a widely quoted SP500 index, if such a product were linked to balanced
bond/stock index, the cost might be different as well, depending on the unique volatility
and growth projections.

That said we[5] have developed a theoretical model for valuing an RCLA assuming a
number of idealized assumptions. These are the same assumptions that are used to
price and hedge widely available equity put options as well as generic life annuities,
which is why we believe the model values provide a rough indication of what an RCLA
would cost in reality.

| |
|---|
| *TABLE #2 PLACED HERE* |

Table #2 provides some estimates of these values assuming a variety of purchase ages
and withdrawal rate index values. Here is how to read the table. For example, a (unisex)
57 year old would have to pay a lump sum $16,667 in exchange for a guarantee of
$6,000 annual inflation-adjusted lifetime income starting if-and-when the vintage 2009

---

[5] Please see the technical working paper by Huang, Milevsky and Salisbury, IFID Centre, Winter 2008, for
a detailed description of the rather subtle issues that arise in attempting to value an RCLA as well as a
sensitivity analysis.

SwP pseudo-index struck at 6%, ever hits zero. The same 57 year old would have to pay $26,983 for $7,000 of real income upon ruin, and only $8,983 for $5,000 of real annual income. Notice that the younger the retiree when they purchase the RCLA the more they must pay (naturally) and the greater the withdrawal rate the more you must pay as well.

Remember that a 7% RCLA is worth much more than a 5% RCLA for two reasons. First of all you are (possibly) getting more income, $7,000 versus $5,000 for example. Second, and more subtly, the underling SwP index upon which the eventual payment is contingent is more likely to hit zero (sooner) since the withdrawals are greater as well.

In the same Table #2 we also display the value (which is the approximate cost) for an immediate annuity which pays an inflation-adjusted lifetime income of $1,000 starting immediately. Thus, for example, at the same age of 57 the retiree would have to pay almost 18 times the amount of desired income. To put this number in perspective, the 7% RCLA would cost $26,983 for $7,000 of (eventual, possible) lifetime income. This is slightly more than 3.8 times the desired income. Compare this to the factor of 18 for the cost of an immediate annuity. The reason for the gap is due to the smaller chance of the RCLA paying out, and at some distant time in the future.

Now, there exists many variations on this basic theme. In theory, one could acquire an RCLA at retirement (i.e. when you start spending) but pay for the insurance protection in installments. For example, instead of paying $26,983 up front at the age of 57 for a 7% RCLA – which is quite steep -- you would be required to pay half now and half if-and-when the SwP pseudo-index level hits zero and the annuity payments start. Alternatively, the buyer could amortize the cost of the RCLA over 5, 10 or 15 years. This would obviously open up the possibility of the buyer lapsing or discontinuing premium payments if the SP500 grew by more than average – which then reduced the chances of the SwP index hitting zero. Another possibility is a super-RCLA that would be based on a periodic step-up in the level of income, based on the performance of the underlying

index. All of these would obviously complicate the pricing and valuation since the optimal dynamic strategy for the retiree would be quite complex.

The bottom line is that we can only provide a rough estimate of what such a product would actually cost, especially in the early stage of the market when few insurance players would offer an RCLA, and likely at inflated prices. Over time of course, as new participants entered the market prices would reach some sort of competitive equilibrium which would resemble the hedging costs, which are closely linked to the model values we just described.

**REMEMBER, A SIMILAR PRODUCT ALREADY EXISTS.**

Astute readers – or at least those who are intimately familiar with the variable annuity industry in the U.S. -- will recognize that a type of RCLA is embedded within each and every guaranteed living benefit (GLiB); as we mentioned in the introduction. For example, recall that a guaranteed withdrawal benefit for life (GMWB, sometime called a GLWB) assures the policyholder that if they withdraw no more than 5% of the initial premium deposit, for example, they will be entitled to receive payments for the rest of their life regardless of the performance of the sub-accounts. And so, if the account value ever hits zero, the guarantee kicks-in and the annuitant receives lifetime payments. With a little bit of imagination one can visualize a GMWB as a combination of a systematic withdrawal plan together with an RCLA. Of course, most GLWB/GMWB products do not offer inflation adjustments to the annual income, on the other hand many do offer periodic step-ups which serve as a type of inflation hedge.

Table #2 can now be interpreted as more than just model values for a theoretical product, but an actual estimate of the discounted value of the embedded longevity and sequence-of-returns insurance offered by a variable annuity with a guaranteed living benefit. A 67 year old who buys a 5% GMWB-for-life is acquiring a guarantee that is worth approximately $3,707 per $100,000 premium deposit. A 57 year old, in contrast, is obtaining a benefit that is worth a much higher $8,983 on a discounted value basis.

Once again, someone who purchased a guaranteed living income benefit (GLiB) is buying a generic investment account – which shouldn't cost anything beyond the money management fees -- together with an RCLA.

The comparative advantage of the insurance industry is protection and risk management, not necessarily investment management and portfolio selection. So, we conclude by asking why not offer the RCLA on its own?

**CONCLUSION AND FURTHER APPLICATIONS**

The creation of a stand-alone ruin contingent life annuity (RCLA) would be a triumph of insurance and financial engineering. On the one hand it is a type of long-term equity put option, but it also provides true longevity insurance. Indeed, it is currently embedded within an assortment of GLiBs on variable annuities, but we believe they should be given a separate life of their own and sold on a stand alone basis. The existence of such a product would benefit even those individuals who opt to buy versions bundled together with other guarantees --- having component prices will help investors evaluate whether the overall fee for a product is reasonable given their circumstances, considering those bundled features they desire, as well as those they are indifferent to.

From a different perspective, a large number of recent papers within the financial planning literature have used advanced mathematical techniques to estimate or forecast the probability that a particular retirement spending rate extracted from a given retirement portfolio, is sustainable or will result in retirement ruin. Ever since the first studies emerged in the early 1990s, various authors have used different techniques, such as historical simulations, Monte Carlo simulations, deterministic formulas and other stochastic approximations to investigate this question. Yet, as many practitioners are now aware, there exists a plethora of recommendations based on assumptions regarding capital market returns, mortality and inflation. At this point, it seems the best the literature can do, is conclude that something between $3 to $6 is sustainable depending on various economic and behavioral assumptions.

Thus, another use of a market for RCLA products would be that it would provide us with a mark-to-market (or at least mark-to-model) value for one's retirement income plan. In other words, if a 7% spending rate is truly unsustainable, then the cost of 7% RCLA would tell us by how much – exactly. In other words, the existence of a liquid market for ruin-contingent life annuities would enable the financial planning industry to refer their clients to actual market prices for sustainable spending rates, as opposed to model values and forecasts.

Stated bluntly, if mom and dad are spending "too much", the relevant x% RCLA value would provide the beneficiaries with a rough (average) estimate of what it will cost them – in discounted value terms – to cover their anticipated spending if and when the retirement account ever runs out. The children might want to set this sum of money aside now, in a risk free saving account, to cover the cost of a life annuity if-and-when mom and dad ever deplete their nest egg, and fall back on their family for support.

Once again, we are offering more than just a product proposal here. The creation of such products would enable retirees and their financial advisor to put a market price – as opposed to just simulation values -- on the risks they are running by spending too much, not investing appropriately or simply living too long. In this case, market prices would function as economic signals or even warning signs.

**Figure #1**

### Evolution of the 7% SwP pseudo-index based on vintage year

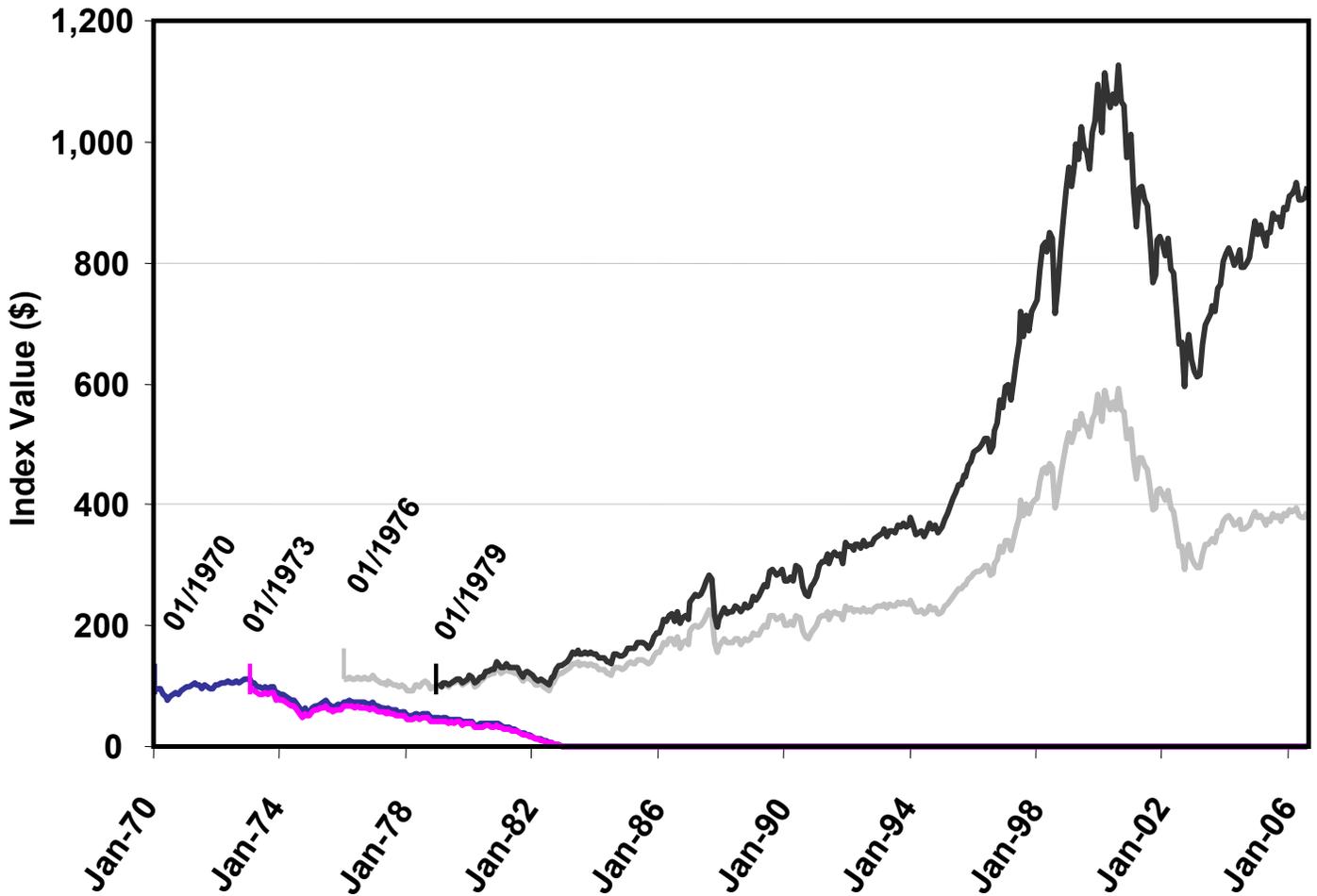

Note: Somebody who retired in 1973, invested their $100 nest egg in the SP500 index and then spent $7 per year adjusted for inflation, would have been ruined in early 1982.

**Table #1**

## The ruin date of the x% SwP pseudo-index:
## Depending on vintage starting year

| | Index initiation date | | | |
|---|---|---|---|---|
| **RCLA Rate** | **Jan 1970** | **Jan 1973** | **Jan 1976** | **Jan 1979** |
| 4% | - | - | - | - |
| 5% | Apr-94 | Oct-90 | - | - |
| 6% | Jan-86 | May-85 | - | - |
| 7% | Jan-83 | Jan-83 | - | - |
| 8% | Jun-81 | Oct-81 | May-03 | - |
| 9% | Feb-80 | Sep-80 | Aug-93 | - |

**Table #2**

## Model Value of a Ruin-Contingent Life Annuity (RCLA)

*Pays inflation-adjusted $100,000 x S% per-year for life*

*if-and-when SwP pseudo-index hits zero…*

| Spending Rate | Purchase Age | | | | |
|---|---|---|---|---|---|
| | **50** | **57** | **62** | **67** | **75** |
| **4.0%** | $6,326 | $3,945 | $2,545 | $1,467 | $440 |
| **5.0%** | $13,687 | $8,983 | $6,072 | $3,707 | $1,256 |
| **6.0%** | $24,410 | $16,667 | $11,680 | $7,459 | $2,779 |
| **7.0%** | $38,292 | $26,983 | $19,469 | $12,891 | $5,192 |

| Value of $1,000 Life Annuity: | $21,838 | $18,810 | $16,493 | $14,102 | $10,304 |
|---|---|---|---|---|---|

Note: A 62 year-old retiree (in 2007, for example) would pay $6,072 up front in exchange for insurance that if the vintage 2007 SwP pseudo-index struck at 5% index ever hit zero, the annuitant would receive an inflation adjusted $5,000 per year for life. Note that the same $5,000 per life annuity starting immediately would cost the 62 year-old $16,493 x 5 = $82,465. The much higher cost of an immediate annuity – promising the same level of income – reflects the fact that the RCLA might never payoff.

*Parameter Assumptions: Risk free rate = 2.5%, expected return (mu) = 7%, volatility of return (sigma) = 20%, both of which correspond with the historical behavior of the SP500 index after-inflation. Mortality, which is used to price the life annuity, is based on the Gompertz m=87.8, b=9.5 model described in Milevsky (2006).*